\documentclass{ifacconf}
\usepackage{amsmath}
\usepackage[ruled,vlined,algo2e]{algorithm2e}
\usepackage{graphicx}      
\usepackage{natbib}        
\usepackage{amsfonts}
\usepackage{lipsum}
\usepackage{pgfplots}
\usetikzlibrary{decorations.pathreplacing}
\usepackage{stfloats}
\pgfplotsset{compat=1.18}
\usepackage[font=small,skip=2pt]{caption}
\begin{document}
\begin{frontmatter}

\title{Addressing Terminal Constraints in Data-Driven Demand Response Scheduling} 


\author[First]{Maximilian Bloor} 
\author[Second]{Martha White} 
\author[First]{Ehecatl Antonio del Rio Chanona}
\author[Third]{Calvin Tsay}

\address[First]{Sargent Centre for Process Systems Engineering, Imperial College London, London, SW7 2AZ, UK}
\address[Second]{Department of Computer Science, University of Alberta, Edmonton, AB, Canada}
\address[Third]{Department of Computing, Imperial College London, London, SW7 2AZ, UK (e-mail: c.tsay@imperial.ac.uk)}

\begin{abstract}                
Electrified chemical processes are incentivized by exposure to time-varying electricity markets to operate flexibly, but participating in demand response schemes can require satisfying terminal constraints over long horizons. 
Specifically, terminal constraints may be required when computing optimal schedules in order to preserve dynamic stability. 
Model-based optimization methods are computationally costly, and data-driven scheduling via reinforcement learning (RL) faces severe credit-assignment challenges. We integrate Goal-Space Planning (GSP) with Deep Deterministic Policy Gradient (DDPG), using learned temporally abstract models over discrete subgoals to propagate value across extended horizons. Using a simulated air separation benchmark, we demonstrate the proposed approach improves sample efficiency over standard DDPG while satisfying terminal storage constraints, mitigating myopic control behavior.
\end{abstract}


\end{frontmatter}

\section{Introduction}

\begin{figure}[h]
	\centering
	\includegraphics[clip, trim=5.2cm 9.1cm 10cm 4.3cm, width=0.5\textwidth]{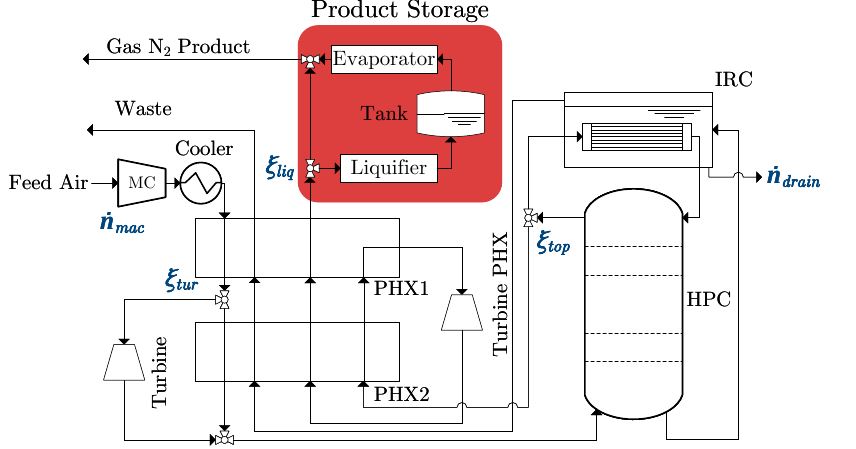}
	
	\caption{ASU Process Flowsheet. Manipulated Variables are marked in blue and the Product Storage Section is shaded in red.}
	\label{fig:ASU_Process_Flowsheet}
\end{figure}

Increasing integration of renewable energy into power grids motivates electrified chemical processes to shift from static consumers to flexible Demand Response (DR) participants~\citep{baldea2025transforming}. Air Separation Units (ASUs), among the largest manufacturing electricity consumers, have inherent temporal flexibility through product storage, making them ideal DR candidates~\citep{pattison_optimal_2016}: ASUs can reduce production during high-price periods and use stored inventory to meet demand (Figure~\ref{fig:ASU_Process_Flowsheet}). Realizing this behavior requires coordinating operations across long horizons while respecting constraints.

Conventional scheduling and control face challenges in DR: first-principle models are difficult to develop and maintain~\citep{tsay2019optimal}, and real-time optimization can become computationally prohibitive~\citep{caspari_integration_2020}. To address this, reduced-order strategies have been proposed, including scale-bridging models~\citep{pattison_optimal_2016} and MPC formulations~\citep{dias_simulation-based_2018, caspari_integration_2020}. Recent work by \citet{schulze2023datadrivenmodelreductionnonlinear} leverages Koopman theory for nonlinear MPC, highlighting the persistent trade-off between complexity and speed.

Reinforcement learning (RL) offers a data-driven alternative with cheap real-time inference since, computation of the optimal policy is shifted offline~\citep{suttonRL}. RL has shown promise in complex chemical processes~\citep{YOO2021108}, with recent work integrating RL with classical controllers~\citep{CIRL,lawrence2022deep} and chance constraints~\citep{PETSAGKOURAKIS202235}. Our recent hierarchical framework~\citep{bloor2025hierarchical} combines a scheduling RL agent with low-level MPC tracking for ASU DR, but struggles with long-horizon credit assignment, e.g., violating terminal storage constraints. More generally, model-free RL suffers in sample efficiency when long horizons are required to satisfy terminal constraints~\citep{bloor2025survey}.

Satisfying terminal constraints is a prominent challenge for RL. Instead of treating this as constrained optimization~\citep{pan2021constrained,burtea2024constrained}, we encode the constraint in the reward function: for the ASU this means incentivizing minimum end-of-horizon storage to ensure stability in subsequent periods, while minimizing operating cost. This complicates credit assignment, since the feedback linking early controls to violations days later yields sparse, delayed signals with weak learning gradients.

In this work, we address the above long-horizon credit assignment problem by reformulating terminal constraints as desirable goal states. Two observations motivate this: (i) $h(\mathbf{x}_T)\leq 0$ depends on a small subset of state components (here, end-of-horizon storage) that discretizes naturally into goals; and (ii) a goal formulation allows distant (in time) value to propagate backwards via temporally abstract dynamic programming, providing a learning signal that hard set-membership penalties cannot. Building on the hierarchical framework of~\cite{bloor2025hierarchical}, we develop Goal-Space Planning (GSP)~\citep{lo2024goalspaceplanning} strategies that equip the RL agent with learned, temporally abstract models for efficient long-horizon value propagation, enabling it to satisfy distant terminal constraints.

\section{Background}
Computing optimal schedules involves solving a dynamic optimization problem over a finite horizon, and deploying solutions in a moving-horizon strategy. 
We formulate the deterministic finite-horizon scheduling problem as a Markov Decision Process (MDP) defined by the tuple $\langle \mathcal{X}, \mathcal{U}, \mathcal{R}, \mathcal{P}, \mathcal{T}\rangle$, with state space $\mathcal{X} \subseteq \mathbb{R}^{n_x}$, control space $\mathcal{U} \subseteq \mathbb{R}^{n_u}$, reward function $\mathcal{R}$, and horizon $\mathcal{T} = \{0, \ldots, T\}$. The transition function $\mathcal{P}$ gives the probability of moving to a new state given the current state and control,
\begin{equation}
    \mathcal{P}(\mathbf{x}_{t+1} \mid \mathbf{x}_t, \mathbf{u}_t) = \prod_{i=1}^{n_x}\delta({x}_{i,t+1} - f_i(\mathbf{x}_t, \mathbf{u}_t)),
\end{equation}
where $\delta(\cdot)$ is the Dirac delta and $\mathbf{f}$ is a discrete-time dynamic model with $\mathbf{x}_{t+1} = \mathbf{f}(\mathbf{x}_t, \mathbf{u}_t)$. At each step, the agent selects control $\mathbf{u}_t \in \mathcal{U}$ and receives a scalar reward $r_t = r(\mathbf{x}_t,\mathbf{u}_t,\mathbf{x}_{t+1})$, seeking a policy $\pi$ that maximizes cumulative reward over the finite horizon.

In RL, the agent typically learns an action-value function $Q(\mathbf{x}_t, \mathbf{u}_t)$ that estimates the expected cumulative reward from taking control $\mathbf{u}_t$ in state $\mathbf{x}_t$. Temporal Difference (TD) learning provides a model-free way of learning this function. The estimate for the current state-action pair $(\mathbf{x}_t, \mathbf{u}_t)$ is updated using the immediate reward and the maximum estimated Q-value of the next state via
\begin{equation}
\begin{split}
Q(\mathbf{x}_t, &\mathbf{u}_t) \leftarrow Q(\mathbf{x}_t, \mathbf{u}_t) +  \\
& \alpha \left[ \left(r_t + \gamma \max_{\mathbf{u}_{t+1}} Q(\mathbf{x}_{t+1}, \mathbf{u}_{t+1})\right) - Q(\mathbf{x}_t, \mathbf{u}_t) \right].
\end{split}
\end{equation}
Here $(r_t + \gamma \max_{\mathbf{u}_{t+1}} Q(\mathbf{x}_{t+1}, \mathbf{u}_{t+1}))$ is the TD target, $\gamma$ the discount factor  (can be set to 1 in the finite horizon setting), and $\alpha$ the learning rate. The update bootstraps by assuming the best possible action is taken from $\mathbf{x}_{t+1}$ onward, so value information propagates backward through time, allowing the agent to assign credit for future outcomes to earlier state-action pairs.

For continuous state-action spaces, Deep Deterministic Policy Gradient (DDPG)~\citep{lillicrap2019continuouscontroldeepreinforcement} is a popular choice of RL algorithm. DDPG is an actor-critic, model-free algorithm that concurrently learns a deterministic policy $\mu(\mathbf{x}_t \mid \theta^\mu)$ (the actor) and a Q-function $Q(\mathbf{x}_t, \mathbf{u}_t \mid \theta^Q)$ (the critic). The actor learns the control policy to be deployed. The critic learns to approximate the above action-value function, and is updated by minimizing the Mean Squared Bellman Error against the TD target
\begin{equation}
y_t = r_t + \gamma Q'(\mathbf{x}_{t+1}, \mu'(\mathbf{x}_{t+1})),
\end{equation}
which is computed using separate target networks $Q'$, $\mu'$ (with parameters $\theta^{Q'}$, $\theta^{\mu'}$) for stability. The resulting critic loss, averaged over a mini-batch of $N$ transitions from a replay buffer, is
\begin{equation}\label{eq:MSBE}
L(\theta^Q) = \frac{1}{N} \sum_{t} \left( y_t - Q(\mathbf{x}_t, \mathbf{u}_t \mid \theta^Q) \right)^2.
\end{equation}
The actor is then updated via the deterministic policy gradient, using the gradient of the critic's Q-function with respect to the actor's actions to push $\theta^\mu$ toward actions the critic predicts will yield higher Q-values. Both the actor and critic are taken as deep neural networks, and the target networks and replay buffer stabilize learning in continuous action spaces.

\section{Problem Formulation}
The DR scheduling problem is cast as a finite-horizon control problem with a hierarchical structure: a high-level RL policy $\pi$ computes a schedule of optimal setpoints $\mathbf{x}^*_t$ over a (longer) scheduling time horizon, and a low-level Linear Model Predictive Controller (LMPC) $\psi$ guides the process to track these setpoints,
\begin{align*}
    \max_{\pi} \quad &\sum^{T-1}_{t=0}r(\mathbf{x}_t, \mathbf{u}_t, \mathbf{x}_{t+1}) \\
    \text{s.t.} \quad &\mathbf{x}_0 = \mathbf{x}(t_0),\quad \mathbf{x}_{t+1} = \mathbf{f}(\mathbf{x}_t, \mathbf{u}_t), \\
    &\mathbf{u}_t = \psi(\mathbf{x}_{t}, \mathbf{x}^*_{t} = \pi(\mathbf{x}_{t})).
\end{align*}
$\psi$ solves an optimization problem over a (shorter) control horizon at each step to track the RL-given setpoints. LMPC is chosen for its computational speed, though its linear approximation can limit performance under large operational changes~\citep{caspari_integration_2020}.
\begin{figure}[h!]
\centering
\scalebox{0.6}{%
\begin{tikzpicture}
    \begin{axis}[
        xlabel={Time},
        ylabel={Goals},
        xmin=0, xmax=4,
        ymin=0, ymax=3,
        grid=major,
        grid style={dashed, gray},
        xtick={0,1,2,3,4},
        ytick={0,1,2,3},
        axis lines=middle,
        xticklabels={},
        yticklabels={},
        tick style={draw=none},
        xlabel style={at=(current axis.right of origin), anchor=west},
        ylabel style={
            at={(rel axis cs:-0.1, 0.65)},
            rotate=90,
            anchor=east,
            xshift=-5pt
        },
        xlabel style={
            at={(rel axis cs:1, -0.1)},
            rotate=0,
            anchor=east,
            xshift=-5pt
        },
        clip=false,
    ]
        \addplot[
            smooth,  
            thick,   
            blue,    
            mark=*,   
            mark options={fill=blue, mark size=1.5pt} 
        ] coordinates {
            (0.2, 0.5)
            (0.5, 1.5) 
            (1.0, 2.5)
            (1.5, 2.2)
            (2.0, 1.8)
            (2.5, 1.5)
            (3.0, 1.9) 
            (3.5, 2.5)
            (3.8, 2.8)
        };
        
        
        \node at (axis cs:0.5, 1.5) [pin=300:{$\mathbf{x}_0$}] {};
        \node at (axis cs:3.0, 1.9) [pin=0:{$\mathbf{x}_1$}] {};
        
        \node at (axis cs:1.5, 0.5) {$\mathbf{g}_0$}; 
        \node at (axis cs:2.5, 2.5) {$\mathbf{g}_1$}; 
        
        \draw [
            decorate,
            decoration={brace, amplitude=5pt, mirror}, 
            thick
        ] 
        (axis cs:1, -0.1) -- (axis cs:2, -0.1) 
        node [midway, below=5pt] {Time Period}; 
    \end{axis}
\end{tikzpicture}
}

\caption{Illustration of goals in the goal-time space.}
\label{fig:storage_goal illustration}
\end{figure}
\begin{figure*}[tbp]
    \centering
    \includegraphics[width=0.75\linewidth]{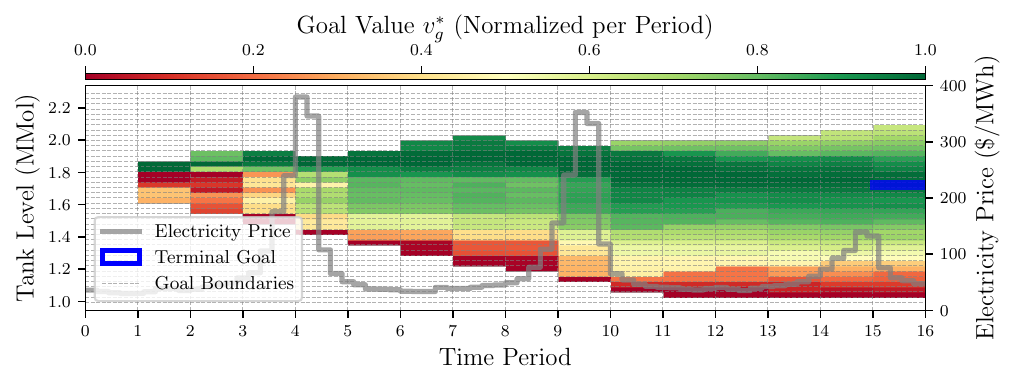}
    
    \caption{Heat map of the learned subgoal values $v_{\mathbf{g}^*}$ using the fully trained goal-to-goal models $\tilde{r}(\mathbf{g},\mathbf{g}')$ and $\tilde{\Gamma}(\mathbf{g},\mathbf{g}')$. The heatmap is normalized over each column. Green represents high values and red represents low values.}
    \label{fig:goal_values}
\end{figure*}
\begin{figure}[h]
    \centering
    \includegraphics[width=0.7\linewidth]{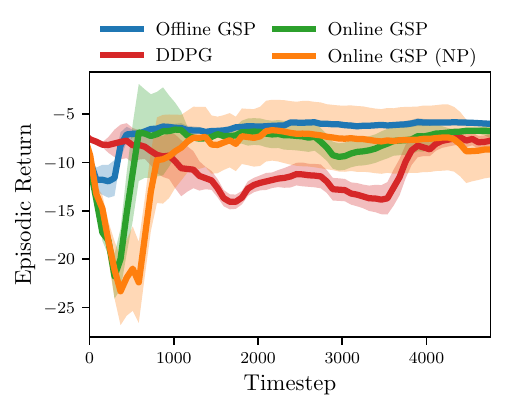}
    
    \caption{Learning curves for DDPG, GSP with models trained offline (Offline GSP), GSP with models trained online (Online GSP), and GSP without the state-to-goal models (Online GSP (NP)). Algorithms were trained for 80 episodes and repeated over 5 seeds (mean and std. shown).}
    \label{fig:learning_curves}
\end{figure}

\subsection{Constraint Handling}\label{sec:constraint}
While constraints can be embedded directly in the DDPG framework~\citep{burtea2024constrained}, we adopt a penalty-based softening that folds operational constraints into the reward. To densify the sparse terminal constraint $h(\mathbf{x}_{T}) \leq 0$, we relax it into the quadratic penalty
\begin{equation} \label{eq:pathcon2}
r_\mathrm{path}(\mathbf{x}_t)= \begin{cases}
\lambda (h(\mathbf{x}_t))^2 & \text{if }h(\mathbf{x}_t) > 0 \text{ and } t > t_a \\
0 & \text{otherwise},
\end{cases}
\end{equation}
with weight $\lambda>0$. The activation time $t_a$ trades the agent's load-shifting freedom against response time, providing continuous feedback over the final stages rather than a single signal at $T$.

\section{Goal-Space Planning}\label{sec:GSP}
Goal-Space Planning (GSP)~\citep{lo2024goalspaceplanning} addresses accumulated errors in long-horizon planning when the problem admits a meaningful definition of subgoal(s). We use GSP as a complement to the penalty in Section~\ref{sec:constraint} to tackle terminal-constraint satisfaction. While \eqref{eq:pathcon2} densifies the signal near $T$, it only heuristically credits controls and can miss the impact of earlier actions. We select GSP over alternatives ($n$-step returns, hindsight replay, hierarchical options, etc.) because its abstract models operate on a low-dimensional, user-defined subgoal space, making dynamic-programming updates cheap and decoupling value propagation from high-frequency dynamics. Learning $\tilde{r}(\mathbf{g},\mathbf{g}')$ and $\tilde{\Gamma}(\mathbf{g},\mathbf{g}')$ between adjacent time periods reduces the effective planning depth from $T$ environment steps to the (much smaller) number of goal-space transitions, so terminal-goal value reaches early states in a step-rate-independent number of Bellman backups.

\subsection{Goal-Space Planning Background}
GSP builds an abstract MDP over a finite subgoal set $\mathcal{G} \in \mathbb{R}^{n_g}$ (Figure~\ref{fig:storage_goal illustration}). For our ASU, we formulate subgoals by quantizing the range of product storage into intervals. The subgoals then summarize all process states into the current storage interval and time window (e.g., 80--90\% full, 12--2 PM), abstracting over high-frequency dynamics like impurities and column temperatures. We denote a membership function $m$ such that state $\mathbf{x}_t$ belongs to subgoal $\mathbf{g}$ if $m(\mathbf{x}_t,\mathbf{g}) = 1$. Likewise we denote a reachability function $d$ such that $\mathbf{g}$ is reachable from $\mathbf{x}_t$ if $d(\mathbf{x}_t,\mathbf{g}) = 1$. Both $m$ and $d$ are user-defined. Subgoal values $\tilde{v}: \mathcal{G} \rightarrow \mathbb{R}$ are computed by value iteration,
\begin{equation}\label{eq:VI}
    \tilde{v}(\mathbf{g}) = \max_{\mathbf{g}'\in {\mathcal{G}}} \left[ \tilde{r}(\mathbf{g},\mathbf{g}') + \tilde{\Gamma}(\mathbf{g},\mathbf{g}')\tilde{v}(\mathbf{g}')\right],
\end{equation}
where $\tilde{r}$ and $\tilde{\Gamma}$ are learned subgoal-to-subgoal models of the discounted return and discount of the transition. State-to-subgoal models $r_{\gamma}(\mathbf{x}_t,\mathbf{g})$ and $\Gamma(\mathbf{x}_t,\mathbf{g})$ similarly estimate the discounted return and discount from state $\mathbf{x}_t$ to $\mathbf{g}$ when $d(\mathbf{x},\mathbf{g})>0$, i.e., the subgoal is reachable.

Abstract values are transferred to the base agent through the potential-based reward-shaping update
\begin{equation}\label{eq:reward_shaping}
    \hat{r}_{t+1} = r_{t+1} + \gamma \Phi(\mathbf{x}_{t+1}) - \Phi(\mathbf{x}_t),
\end{equation}
which adds a potential difference to the reward without affecting the optimal policy~\citep{ng1999policy}. We define the potential function as the projected subgoal value
\begin{equation}\label{eq:v_g^star}
    v_{g^*}(\mathbf{x}_t) = \max_{\mathbf{g} \in \mathcal{\hat{G}}} \left[r_\gamma (\mathbf{x}_t,\mathbf{g}) + \Gamma(\mathbf{x}_t,\mathbf{g})\tilde{v}(\mathbf{g})\right],
\end{equation}
which estimates the value of $\mathbf{x}_t$ by combining the learned state-to-subgoal models with the subgoal values $\tilde{v}(\mathbf{g})$ and selecting the most valuable reachable nearby subgoal. 
Here $\mathcal{\hat{G}}$ is the set of `nearby' subgoals, defined using a prespecified distance metric in goal space. 
Some example subgoal values are illustrated in Figure~\ref{fig:goal_values}, where the subgoals are tank levels reachable at various time periods. 
Substituting the shaped reward gives the DDPG critic target
\begin{equation}
    (r_{t+1} + \gamma v_{g^*}(\mathbf{x}_{t+1}) - v_{g^*}(\mathbf{x}_t)) + \gamma Q'(\mathbf{x}_{t+1}, \mu'(\mathbf{x}_{t+1})).
\end{equation}
This shapes immediate feedback by rewarding moves toward higher-value subgoals (i.e., $v_{g^*}(\mathbf{x}_{t+1}) - v_{g^*}(\mathbf{x}_t) > 0$) and penalizing the opposite. Because $v_{g^*}$ abstracts states across time, it is approximate but propagates value quickly through the horizon, enabling the agent to learn the long-horizon control needed for terminal-constraint satisfaction.

\subsection{GSP Implementation}
The state-to-subgoal models $r_{\gamma}(\mathbf{x}_t,\mathbf{g})$ and $\Gamma(\mathbf{x}_t,\mathbf{g})$ above are parametrized as neural networks, which we denote as $r^\omega_{\gamma}$ and $\Gamma^\eta$ with parameters $\omega$ and $\eta$, respectively. 
Moreover, the two are implemented as separate heads of a shared body network.  
Their training dataset $\mathcal{D}$ (collected offline or online) records valid state-to-goal transitions with discounted return $r_\gamma = \sum_{k=0}^{h-1} \gamma^k r_{t+k}$ and discount $\Gamma = \gamma^h$, where $h$ is the source-to-target step count. Inputs concatenate state and goal coordinates. The combined loss is
\begin{multline}\label{eq:gsp_models}
    \mathcal{L}(\omega, \eta) = \frac{1}{|\mathcal{D}|} \sum_{(\mathbf{x}_t, \mathbf{g}, r_\gamma, \Gamma) \in \mathcal{D}} \left[ \left( r^\omega_{\gamma}(\mathbf{x}_t, \mathbf{g}) - r_\gamma \right)^2 \right. \\
    \left. + \left( \Gamma^\eta(\mathbf{x}_t, \mathbf{g}) - \Gamma \right)^2 \right].
\end{multline}

The goal-to-goal models $\tilde{r}_\gamma$ and $\tilde{\Gamma}$ are built directly from $\mathcal{D}$ without parametric approximation, since $\mathcal{G}$ is suitably small. For each episode, transitions between goals are identified when the agent moves from one time period to the next, and the discounted cumulative reward $\tilde{r}_\gamma(\mathbf{g}, \mathbf{g}')$ and discount $\tilde{\Gamma}(\mathbf{g}, \mathbf{g}')$ are computed by aggregating rewards over the period. These transitions are stored in a directed acyclic graph (DAG) with edges that represent feasible goal-to-goal transitions, and edge attributes storing the mean $\tilde{r}_\gamma$ and $\tilde{\Gamma}$ across all observed instances. Integration with DDPG is summarized in Algorithm~\ref{alg:ddpg-gsp}.

\begin{algorithm2e}[h]
\caption{DDPG with Online GSP}
\label{alg:ddpg-gsp}
\SetAlgoLined
\DontPrintSemicolon
\footnotesize
Init. actor $\mu$, critic $Q$, targets, buffer $\mathcal{D}$, subgoals $\mathcal{G}$, values $\tilde{v}$\;
\For{episode $e \leftarrow 1$ \KwTo $M$}{
    \For{$t \leftarrow 0$ \KwTo $T-1$}{
        Select action $\mathbf{u}_t = \mu(\mathbf{x}_t) + \mathcal{N}(0,\sigma) $\;
        Observe $\mathbf{x}_{t+1} = \mathbf{f}(\mathbf{x}_t, \mathbf{u}_t)$ and $r_t = r(\mathbf{x}_t, \mathbf{u}_t, \mathbf{x}_{t+1})$\;
        Store $(\mathbf{x}_t, \mathbf{u}_t, r_t, \mathbf{x}_{t+1})$ in $\mathcal{D}$\;
        \If{training}{
            Sample batch; Compute potentials $v_{g^*}$ (Eq. \ref{eq:v_g^star})\;
            Shape reward: $\hat{r}_t = r_t + \gamma \Phi(\mathbf{x}_{t+1}) - \Phi(\mathbf{x}_t)$\;
            Update $\theta^Q, \theta^\mu$ and targets $\theta^{Q'}, \theta^{\mu'}$ using $\hat{r}_t$ (Eq. \ref{eq:MSBE})\;
        }
    }
    Update GSP: Construct \& prune graph from $\mathcal{D}$\;
    Train models $\omega, \eta$ (Eq. \ref{eq:gsp_models}) and run VI (Eq. \ref{eq:VI}) to update $\tilde{v}$\;
}
\Return{$\theta^\mu$}
\end{algorithm2e}

Building the graph directly from observed trajectories often introduces connectivity artifacts, such as premature terminal goals that fail to reach the final time period or unnatural starting goals disconnected from the initial period. We apply a two-stage pruning mechanism to enforce full connectivity. Let $\mathcal{V}_T$ denote the goals in the terminal time period and $\mathcal{V}_0$ those reachable from the initial state. \emph{Backward pruning} performs a reverse breadth-first search from $\mathcal{V}_T$ along the edges in reverse and removes any node not visited, eliminating dead ends that cannot reach the terminal set. \emph{Forward pruning} then performs a forward breadth-first search from $\mathcal{V}_0$ and removes any node not visited, discarding goals unreachable from the initial conditions. The resulting subgraph contains only goals on at least one complete $\mathcal{V}_0\!\to\!\mathcal{V}_T$ trajectory, preventing disconnected nodes from distorting~\eqref{eq:VI}.

\section{Computational Case Study}

We use the simulated benchmark process ASU of \citet{tsay2020benchmark} (Figure~\ref{fig:ASU_Process_Flowsheet}), which produces high-purity nitrogen via a single cryogenic distillation column. Flexibility comes from liquefying product into storage, which can later be vaporized to meet product demand, enabling load shifting behavior. The liquefied fraction
\begin{align}
	\xi_{\text{liq}} = \begin{cases}
		1 - \frac{\dot{n}_{\text{demand}}}{\dot{n}_{\text{product}}} & \text{if } \dot{n}_{\text{product}} > \dot{n}_{\text{demand}} \\
		0 & \text{otherwise}
	\end{cases}
\end{align}
is chosen so that production plus evaporation meets demand at all times.
We refer the reader to \citet{caspari_integration_2020, pattison_optimal_2016} for full details regarding the ASU model and its computational implementation.

The agent observes the 17-dimensional state vector
\begin{equation*}
\mathbf{x}_t = \{I_\mathrm{product},\, \Delta T_\mathrm{IRC},\, N_\mathrm{tank},\, F_\mathrm{tank},\, \hat{p}_t,\ldots, \hat{p}_{t+11},\, t_d\},
\end{equation*}
consisting of product impurity $I_\mathrm{product}$, the reboiler-condenser temperature gap $\Delta T_\mathrm{IRC}$, storage holdup $N_\mathrm{tank}$, inflow to storage $F_\mathrm{tank}$, a 12-hour price forecast, and time-of-day $t_d$. It outputs a continuous product setpoint $\dot{n}_\mathrm{product}^*$, which the lower-level MPC of \citet{dias_simulation-based_2018} tracks via four inputs: air compressor feed $\dot{n}_\mathrm{mac}$, recycle fraction $r_\mathrm{gas}$, primary heat-exchanger bypass $\xi_\mathrm{PHX}$, and high-pressure column drain $\dot{n}_\mathrm{drain}$. The reward is
\begin{equation}
    r_t = r_{t,\mathrm{elec}} + r_{t,\mathrm{path}} + r_{t,\mathrm{terminal}},
\end{equation}
with $r_{t,\mathrm{elec}} = - p_t(P_\mathrm{comp}+P_\mathrm{liq}-P_\mathrm{tur})\Delta t$ the total electricity cost, $r_{t,\mathrm{path}}$ active in the final 4 hours per~\eqref{eq:pathcon2}, and $r_{t,\mathrm{terminal}}$ rewarding meeting the terminal constraint, i.e., final storage within tolerance of the target.

\subsection{GSP Formulation for the ASU}

Subgoals are defined over storage level and time, encoding two pieces of domain knowledge: (i) the terminal constraint depends only on end-of-horizon storage, so storage is the natural abstraction; and (ii) prices vary hourly, so time-of-day must be retained to align value propagation with price signals. Faster states (column temperatures, impurities) are left to the LMPC. Following this strategy, we formulate the goal space $\mathcal{G}$ as a grid of $40$ storage levels and $16$ time periods over the 72-hour episode ($\sim$4.5 h each), giving $600$ goals (the initial period is excluded as its initiation set is empty). Each subgoal $\mathbf{g}$ is indexed by period $q\in\{1,\ldots,16\}$ and level $\ell\in\{1,\ldots,40\}$. The level spacing ($\sim$2.5\% of capacity) is finer than the terminal tolerance, and the period count is the smallest that resolves day-night oscillations. Coarser grids degraded constraint satisfaction in pilot tests, while finer grids required more data to populate $\tilde{r}$ and $\tilde{\Gamma}$ without significant performance gain. A state belongs to a goal if its storage is within tolerance of the level and its time within the period window.

\subsection{Implementation Details}
The DDPG agent is implemented on top of the Stable Baselines 3 (SB3) library~\citep{stable-baselines3}, with the GSP reward shaping integrated directly into the SB3 training loop as described in Section~\ref{sec:GSP}. Both the actor and critic are deep neural networks trained with Adam, and target networks are updated by Polyak averaging. The state-to-goal networks use a shared two-layer body with separate heads for $r^\omega_\gamma$ and $\Gamma^\eta$, retrained at the end of each episode in the online setting. Goal-to-goal tables and the DAG are likewise rebuilt and pruned per episode before value iteration. The hyperparameter settings for the underlying DDPG agent are listed in Table~\ref{tab:ddpg_hyperparams}, and the same hyperparameter values across all algorithms compared in Figure~\ref{fig:learning_curves} to isolate the effect of GSP.

\begin{table}[h]
\centering
\caption{DDPG hyperparameters used in all experiments.}
\label{tab:ddpg_hyperparams}
\begin{tabular}{ll}
\hline
Hyperparameter & Value \\
\hline
Actor learning rate & $3\times10^{-4}$ \\
Critic learning rate & $3\times10^{-4}$ \\
Replay buffer size & 50{,}000 \\
Batch size & 256 \\
Discount factor ($\gamma$) & 0.99 \\
Target update ($\tau$) & 0.005 \\
Action noise (std.\ dev.) & 0.1 \\
Learning starts (steps) & 1000 \\
\hline
\end{tabular}
\end{table}


\subsection{Results}

Training spans 80 episodes of 72 hourly step each. 
We consider a fixed 72-hour interval of electricity prices proposed by the benchmark process~\citep{caspari_integration_2020}. 
Figure~\ref{fig:learning_curves} compares the learning curves of standard DDPG and various GSP formulations. The results show GSP substantially improves sample efficiency over standard DDPG~\citep{stable-baselines3}: `Online GSP' nears optimal by step 1000, and `Offline GSP' (pre-trained models) converges within 500. The intermediate `Online GSP (NP)' ablation confirms that state-to-goal projection supplies essential credit-assignment information. Overhead is modest: Online GSP averages $\sim$15\% longer wall-clock per episode than DDPG (graph rebuild, VI, and model updates), while Offline GSP adds only the per-step potential evaluation ($<$5\%).

\begin{figure}[tbp]
    \centering
    \includegraphics[width=0.7\linewidth]{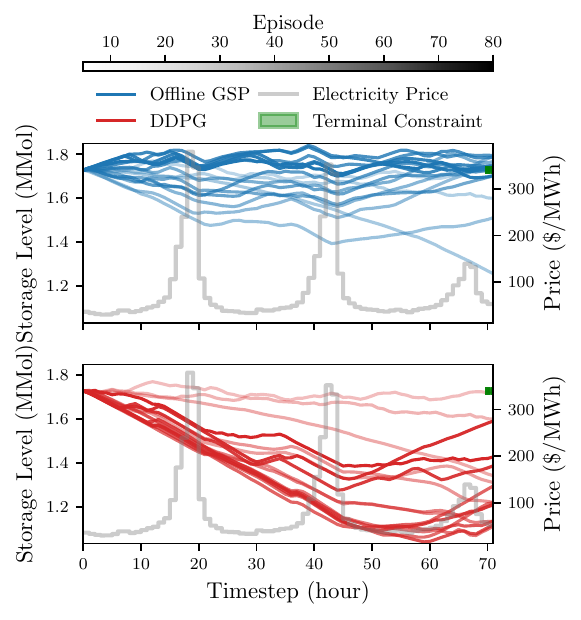}
    
    \caption{Storage level trajectories during training (Episodes 5--80). The opacity of the trajectories indicates the progress through training. The top plot shows the trajectories from the Offline GSP algorithm and the bottom displays them from the DDPG algorithm.}
    \label{fig:storage_trajectories}
\end{figure}

Storage trajectories incurred using the policies obtained during training of the various algorithms are shown in Figure~\ref{fig:storage_trajectories}, revealing behavioral differences. DDPG produces a myopic policy, depleting storage for immediate cost reduction and violating the terminal constraint. Policies obtained using GSP maintain elevated storage and strategically respond to prices (opacity indicates training progress) while satisfying the constraint.

\begin{figure*}
    \centering
    \includegraphics[width=0.75\linewidth]{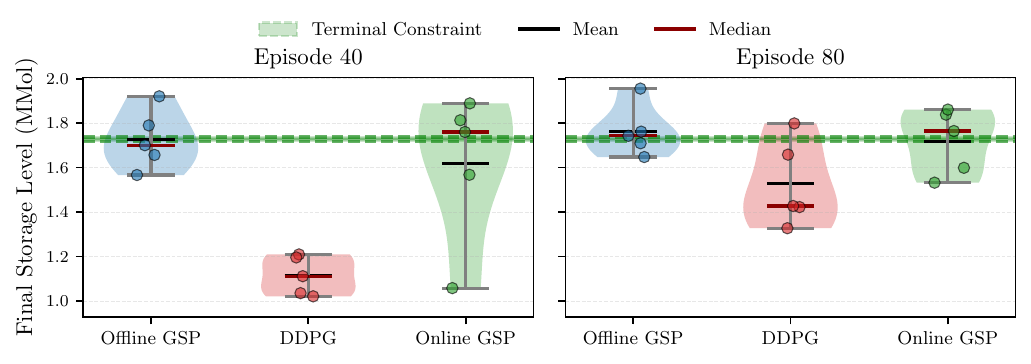}
    
    \caption{Distribution of final storage levels at episodes 40 and 80 for Offline GSP, Online GSP, and DDPG across 5 seeds. The green zone represents the required terminal constraint. Violin plots show distribution density with overlaid individual points. Black and red lines indicate mean and median values.}
    \label{fig:terminal_constraints}
\end{figure*}

Figure~\ref{fig:terminal_constraints} tracks terminal-constraint satisfaction by the various obtained policies. At episode 40, the DDPG policy undershoots the terminal storage level significantly, consistent with the myopic depletion in Figure~\ref{fig:storage_trajectories}, from which recovery is difficult. Online GSP is more variable but produces policies that maintain storage near the target, reflecting subgoal-informed planning (Figure~\ref{fig:goal_values}). By episode 80 all methods generally reflect constraint satisfaction.

The subgoal-value heatmap in Figure~\ref{fig:goal_values} (from applying~\eqref{eq:VI} to the trained $\tilde{r}$ and $\tilde{\Gamma}$) reveals GSP's planning mechanism: high values for elevated early-period storage and increasingly negative values for low storage near termination, encoding the intuitive long-term consequences that counteract myopic tendencies during early training.

The gap between Online GSP and Online GSP (NP) further isolates the contribution of the state-to-goal projection. Without it, the agent only sees abstract goal-to-goal values aggregated across time and cannot resolve which states within a period are progressing toward a valuable terminal goal. With the projection in place, the shaped reward in~\eqref{eq:reward_shaping} becomes a fine-grained, per-step signal aligned with the long-horizon value field of Figure~\ref{fig:goal_values}. Offline GSP converges fastest because its goal-space models are already representative of the optimal value structure at the start of training, whereas Online GSP must build the DAG and refine $\tilde{r}$ and $\tilde{\Gamma}$ from agent experience. The online variant nevertheless has the practical advantage that it does not require an offline data-collection phase and adapts naturally if the price profile or demand changes during training.

\section{Conclusion}
This work demonstrates that Goal-Space Planning (GSP) can effectively address the long-horizon credit assignment problem of satisfying terminal constraints in reinforcement learning (RL)-based demand response scheduling. Using an air separation case study, we show that, by integrating temporally abstract models, GSP enables model-free RL agents to establish credit assignment between early control actions and distant terminal outcomes, reducing convergence time by $\sim$5000 steps compared to standard DDPG, while also more consistently satisfying the terminal storage constraint. Critically, even when trained fully online, the state-to-goal projection supplies the fine-grained temporal information that prevents the myopic control behavior observed in policies from standard DDPG. Examination of the learned goal value functions shows that GSP agents acquire interpretable representations of temporal trade-offs, demonstrating that constructing goal spaces aligned with key constraint variables is an effective mechanism for injecting domain knowledge into model-free RL. These results suggest that goal-based abstractions are a promising route for tackling other long-horizon process-systems control problems in which terminal or path constraints currently limit the applicability of data-driven methods.

\begin{ack}
MB would like to acknowledge funding provided by the EPSRC, UK through grant code EP/W524323/1. CT gratefully acknowledges funding from a BASF/Royal Academy of Engineering Senior Research Fellowship. The authors would also like to thank Haseeb Shah, Jiamin He, and Parham Panahi for insightful discussions.
\end{ack}

\bibliography{ifacconf}             

@article{lo2024goalspaceplanning,
  title={Goal-space planning with subgoal models},
  author={Lo, Chunlok and Roice, Kevin and Panahi, Parham Mohammad and Jordan, Scott M and White, Adam and Mihucz, Gabor and Aminmansour, Farzane and White, Martha},
  journal={J. Mach. Learn. Res.},
  volume={25},
  number={330},
  pages={1--57},
  year={2024}
}

@inproceedings{ng1999policy,
  title={Policy invariance under reward transformations: Theory and application to reward shaping},
  author={Ng, Andrew Y and Harada, Daishi and Russell, Stuart},
  booktitle={ICML},
  volume={99},
  pages={278--287},
  year={1999},
  organization={PMLR}
}

@article{pattison_optimal_2016,
	title = {Optimal Process Operations in Fast-Changing Electricity Markets: Framework for Scheduling with Low-Order Dynamic Models and an Air Separation Application},
	volume = {55},
	language = {en},
	number = {16},
	urldate = {2024-10-08},
	journal = {Ind. Eng. Chem. Res.},
	author = {Pattison, Richard C. and Touretzky, Cara R. and Johansson, Ted and Harjunkoski, Iiro and Baldea, Michael},
	month = apr,
	year = {2016},
	pages = {4562--4584},
}

@article{caspari_integration_2020,
	title = {The integration of scheduling and control: {Top}-down vs. bottom-up},
	volume = {91},
	issn = {09591524},
	language = {en},
	urldate = {2024-08-01},
	journal = {J. Process Control},
	author = {Caspari, Adrian and Tsay, Calvin and Mhamdi, Adel and Baldea, Michael and Mitsos, Alexander},
	year = {2020},
	pages = {50--62},
}

@article{dias_simulation-based_2018,
	title = {A simulation-based optimization framework for integrating scheduling and model predictive control, and its application to air separation units},
	volume = {113},
	issn = {00981354},
	language = {en},
	urldate = {2024-10-02},
	journal = {Comput. Chem. Eng.},
	author = {Dias, Lisia S. and Pattison, Richard C. and Tsay, Calvin and Baldea, Michael and Ierapetritou, Marianthi G.},
	month = may,
	year = {2018},
	pages = {139--151},
}

@article{lillicrap2019continuouscontroldeepreinforcement,
      title={Continuous control with deep reinforcement learning}, 
      author={Timothy P. Lillicrap and Jonathan J. Hunt and Alexander Pritzel and Nicolas Heess and Tom Erez and Yuval Tassa and David Silver and Daan Wierstra},
      year={2019},
      journal={arXiv:1509.02971}
}

@book{suttonRL,
author = {Sutton, Richard S. and Barto, Andrew G.},
title = {Reinforcement Learning: An Introduction},
year = {2018},
isbn = {0262039249},
publisher = {A Bradford Book},
address = {Cambridge, MA, USA},
}

@article{PETSAGKOURAKIS202235,
title = {Chance constrained policy optimization for process control and optimization},
journal = {J. Process Control},
volume = {111},
pages = {35-45},
year = {2022},
issn = {0959-1524},
author = {Panagiotis Petsagkourakis and Ilya Orson Sandoval and Eric Bradford and Federico Galvanin and Dongda Zhang and Ehecatl Antonio del Rio-Chanona},

}

@article{lawrence2022deep,
  title={Deep reinforcement learning with shallow controllers: An experimental application to {PID} tuning},
  author={Lawrence, Nathan P and Forbes, Michael G and Loewen, Philip D and McClement, Daniel G and Backstr{\"o}m, Johan U and Gopaluni, R Bhushan},
  journal={Control Eng. Pract.},
  volume={121},
  pages={105046},
  year={2022},
  publisher={Elsevier}
}

@article{CIRL,
  title={Control-informed reinforcement learning for chemical processes},
  author={Bloor, Maximilian and Ahmed, Akhil and Kotecha, Niki and Mercangoz, Mehmet and Tsay, Calvin and del Rio-Chanona, Ehecatl Antonio},
  journal={Ind. Eng. Chem. Res.},
  volume={64},
  number={9},
  pages={4966--4978},
  year={2025},
  publisher={ACS Publications}
}

@article{YOO2021108,
title = {Reinforcement learning for batch process control: Review and perspectives},
journal = {Annu. Rev. Control},
volume = {52},
pages = {108-119},
year = {2021},
issn = {1367-5788},
author = {Haeun Yoo and Ha Eun Byun and Dongho Han and Jay H. Lee},
}

@article{tsay2019optimal,
  title={Optimal demand response scheduling of an industrial air separation unit using data-driven dynamic models},
  author={Tsay, Calvin and Kumar, Ankur and Flores-Cerrillo, Jesus and Baldea, Michael},
  journal={Comput. Chem. Eng.},
  volume={126},
  pages={22--34},
  year={2019},
  publisher={Elsevier}
}

@article{bloor2025hierarchical,
 title={Hierarchical {RL-MPC} for demand response scheduling},
  author={Bloor, Maximilian and Chanona, Ehecatl Antonio Del Rio and Tsay, Calvin},
  journal={IFAC-PapersOnLine},
  volume={59},
  number={6},
  pages={229--234},
  year={2025},
  publisher={Elsevier}
}

@article{pan2021constrained,
  title={Constrained model-free reinforcement learning for process optimization},
  author={Pan, Elton and Petsagkourakis, Panagiotis and Mowbray, Max and Zhang, Dongda and del Rio-Chanona, Ehecatl Antonio},
  journal={Comput. Chem. Eng.},
  volume={154},
  pages={107462},
  year={2021},
  publisher={Elsevier}
}

@article{baldea2025transforming,
  title={Transforming the Process Industries through Electrification: Challenges and Opportunities},
  author={Baldea, Michael and Endler, Elizabeth E and Hale, Elaine and Maravelias, Christos T and Barolo, Massimiliano and Harjunkoski, Iiro and Mercangoz, Mehmet and Shah, Sirish L and Soroush, Masoud and Young, Brent R and others},
  journal={Ind. Eng. Chem. Res.},
  volume={64},
  number={34},
  pages={16466--16478},
  year={2025},
  publisher={ACS Publications}
}

@article{bloor2025survey,
  title={A survey and tutorial of reinforcement learning methods in process systems engineering},
  author={Bloor, Maximilian and Mowbray, Max and del Rio Chanona, Ehecatl Antonio and Tsay, Calvin},
  journal={Comput. Chem. Eng.},
  pages={109515},
  year={2025},
  publisher={Elsevier}
}

@article{tsay2020benchmark,
  title={A benchmark air separation unit for process control and flexible operation},
  author={Tsay, C and Caspari, A and Pattison, R and Johansson, T and Mitsos, A and Baldea, M},
  journal={Mendeley Data v1},
  volume={2},
  year={2020},
  note={doi:10.17632/pfcc5gvzty.1}
}

@article{schulze2023datadrivenmodelreductionnonlinear,
  title={Data-Driven Model Reduction and Nonlinear Model Predictive Control of an Air Separation Unit by Applied {Koopman} Theory},
  author={Schulze, Jan C and Doncevic, Danimir T and Erwes, Nils and Mitsos, Alexander},
  journal={arXiv:2309.05386},
  year={2023}
}

@article{stable-baselines3,
  author  = {Antonin Raffin and Ashley Hill and Adam Gleave and Anssi Kanervisto and Maximilian Ernestus and Noah Dormann},
  title   = {{Stable-Baselines3}: Reliable Reinforcement Learning Implementations},
  journal = {J. Mach. Learn. Res.},
  year    = {2021},
  volume  = {22},
  number  = {268},
  pages   = {1-8}
}

@article{burtea2024constrained,
  title={Constrained continuous-action reinforcement learning for supply chain inventory management},
  author={Burtea, Radu and Tsay, Calvin},
  journal={Comput. Chem. Eng.},
  volume={181},
  pages={108518},
  year={2024},
  publisher={Elsevier}
}

\end{document}